\documentclass[fleqn,twoside,a4paper]{article}

\usepackage{%
  espcrc2, axodraw, amsmath, amssymb, subfig, url, cite, float,
  graphicx } \mathindent=0pt
\usepackage[figuresright]{rotating}

 \newskip\humongous \humongous=0pt plus 1000pt minus
100pt \def\caja{\mathsurround=0pt}
\def\eqalign#1{\,\vcenter{\openup1\jot \caja \ialign{\strut
      \hfil$\displaystyle{##}$&$
      \displaystyle{{}##}$\hfil\crcr#1\crcr}}\,} \newif\ifdtup

\newcommand{\cf}{\textit{cf.}}

 \newcommand{\Tau}{\tau}

\DeclareMathOperator*{\Res}{Res} \def\Rec{R} \def\la {\langle}
\def\ra{\rangle} 
\def\spa#1.#2{\left\langle#1\,#2\right\rangle}
\def\spb#1.#2{\left[#1\,#2\right]}
\def\spba#1.#2.#3{\left[#1|#2|#3\right\rangle}
\def\spab#1.#2.#3{\left\langle#1|#2|#3\right]}
\def\spaa#1.#2.#3{\left\langle#1|#2|#3\right\rangle}
\def\spbb#1.#2.#3{\left[#1|#2|#3\right]}
\def\lor#1.#2{\left(#1\,#2\right)}  
 \def\oneloop{\text{1-loop}}
 \def\tree{\text{tree}}

\newlength{\colboxsep} 
\newenvironment{colbox}%
{\begin{minipage}{\columnwidth}\vspace\colboxsep}
  {\vspace\colboxsep\end{minipage}}

\title{Augmented Recursion For One-loop Amplitudes}

\author{%
  David~C.~Dunbar, James~H.~Ettle and Warren~B.~Perkins\\
  Department of Physics,\\
  Swansea University,\\
  Swansea, SA2 8PP, UK\\
}

\begin{document}

\begin{abstract}
  We present a semi-recursive method for calculating the rational
  parts of one-loop amplitudes when recursion produces double poles.
  We illustrate this with the graviton scattering amplitude
  $M^\oneloop(1^-,2^+,3^+,4^+,5^+)$.  \vspace{1pc}
\end{abstract}

\maketitle

\section{Introduction}

On-shell recursive techniques, using the rationality of tree
amplitudes and their complex factorisation properties, have proven
very successful in the computation of scattering amplitudes in gauge
and gravity theories~\cite{Britto:new,BBSTgravity}.  Specifically, in
a theory with massless states, if we use a spinor helicity
representation for the polarisation vectors it is possible to write
the amplitude entirely in terms of spinorial variables
$A(\lambda^i_\alpha,\bar\lambda^i_{\dot \alpha})$ where the massless
momentum of the $i^\text{th}$ particle is
$\lambda_\alpha^i\bar\lambda_{\dot \alpha}^i=(\sigma_\mu)_{\alpha\dot
  \alpha} k^\mu_i$.  The analytic structure of the amplitude can be
probed by choosing a pair $a,b$ of external momenta and shifting these
according to
\begin{equation}
  \bar\lambda^a \longrightarrow \bar\lambda^a-z\bar\lambda^b,\quad 
  \lambda^b \longrightarrow     \lambda^b+z    \lambda^a 
  \label{eqshift}
\end{equation}
where we suppress the spinor indices.  If the shifted amplitude $A(z)$
(a) is a rational function, (b) has finite order poles at points $z_i$,
and (c) vanishes as $z\longrightarrow \infty$, then applying Cauchy's
theorem to $A(z)/z$ with a contour at infinity yields
\begin{equation}
  A(0)=-\sum_{\text{poles $z_i$}} \Res_{z=z_i}
      {A(z)\over z}.
  \label{EqCauchyRes}
\end{equation}

At tree level the factorisation of amplitudes is simple: amplitudes
must factorise on multi-particle and collinear poles into the product
of two tree amplitudes defined at $z=z_i$. Thus we can express the
$n$-point tree amplitude in terms of lower point amplitudes,
\begin{equation}
  A_n^\tree (0) \; = \; \sum_{i,\sigma} {A^{\tree,\sigma}_{r_i+1}(z_i)
    {i\over K^2}A^{\tree,-\sigma}_{n-r_i+1}(z_i)},
  \label{RecursionTree}
\end{equation}
where the summation over $i$ is only over factorisations where the $a$
and $b$ legs are on opposite sides of the pole.  This technique is
very effective in computing tree amplitudes and has been extended to a
variety of other applications including 
gravity~\cite{BBSTgravity}.

Beyond tree level there are three potential barriers to using
recursion. Firstly, the amplitudes generally contain non-rational
functions such as logarithms and dilogarithms; secondly, the
amplitudes may contain higher-order poles for complex momenta; and
finally, the amplitudes may not vanish asymptotically with $z$.
Nonetheless a variety of techniques based upon recursion and unitarity
have been developed.  A one-loop amplitude for massless particles may
be expressed as
\begin{equation}
  A^{\oneloop} = \sum_{n=2,3,4;i} c_i I^i_{n} +R, 
\end{equation}
where the scalar integral functions $I^i_{n}$ are the various scalar
box, triangle and bubble functions.  The amplitude can thus be
determined by computing the rational coefficients, $c_i$, and the
purely rational term $R$.  The $c_i$ can be computed by the
four-dimensional unitarity
technique~\cite{BDDKa,Britto:2004nc,Dunbar:2009ax} or indeed
recursively~\cite{Bern:2005hh}.  Many techniques have been developed
for evaluating $R$: $D$-dimensional unitarity, recursion and
specialised Feynman diagram techniques~\cite{Berger:2006ci,
  DunitarityA, DunitarityB, DunitarityC, DunitarityD, DunitarityE,
  DunitarityF, blackhat, Britto:2004ap, Xiao:2006vt, Binoth:2006hk,
  Badger:2008cm, OPP, Bern:1996ja}.

In general, the rational term $R$ does not simply satisfy the
requisites for recursion.  If the amplitude has only simple poles but
does not vanish as $z\longrightarrow \infty$ then it can be possible
to formulate auxiliary recursion relations~\cite{Berger:2006cz}.
However there are rational amplitudes for which one cannot find a
shift which only generates simple poles such as the single-minus
amplitudes $A^\oneloop(1^-,2^+,\cdots,n^+)$.  These amplitudes vanish
at tree level and consequently are purely rational at one-loop.  A
shift on these amplitudes yields double and single poles.
The double pole is {\it not} in itself a a barrier to using recursion,
however to obtain the full residue one needs to know the coincident
single pole, or the `pole under the double pole', which is not
determined by factorisation into on-shell amplitudes.  In
\cite{Bern:2005hs} for Yang--Mills this was postulated to be
\begin{equation}
\frac 1{(K^2)^2} +
\frac{S(a_1,\hat K^+,a_2) S(b_1,\hat K^-, b_2)}{K^2},
\label{eqYangMillsSoft}
\end{equation}
where the `soft' factors are
 $S(a, s^+, b)= {\spa{a}.{b} /
 ( \spa{a}.s\spa{s}.b ) }$,
 $S(a, s^-, b)= {\spb{a}.{b} /
 ( \spb{a}.s\spb{a}.b ) }$,
and $a_1$, $a_2$ ($b_1$, $b_2$) are colour-adjacent to $K$ on the
left (right) side of the pole.
With this ansatz, recursion correctly
reproduces the known single-minus one-loop amplitudes.  In
\cite{Vaman:2008rr} it was shown that the consistency requirements for
recursion in QCD are sufficient to determine these soft factors.

The above postulate, or variations thereof, does not work for gravity
amplitudes~\cite{Brandhuber:2007up}.  Here, we apply a semi-recursive
technique for gravity scattering amplitudes that obtains the `pole
under the pole' using an axial gauge formalism to calculate the
previously-unknown amplitude $M^\oneloop(1^-,2^+,3^+,4^+,5^+)$.  We
assume that the shifted amplitudes vanish as
$z\longrightarrow\infty$.  The derived amplitude has the correct
symmetries and soft limits, providing strong evidence for the validity
of this assumption.  Further, we have checked the result by a
completely independent `string-based
rules'~\cite{Bern:1991aq,Bern:1993wt} computation.

\section{Recursion}
The factorisation of one-loop massless amplitudes is described in
\cite{BernChalmers},
\begin{multline}
  \label{eq:LoopFact}
  A_{n}^{\oneloop} \mathop{\longrightarrow} \sum \biggl[
  A_{r+1}^{\oneloop}\, {i \over K^2} \, A_{n-r+1}^{\tree}
  \\
  + A_{r+1}^{\tree} {i\over K^2} A_{n-r+1}^{\oneloop} +
  A_{r+1}^{\tree} {i\over K^2} A_{n-r+1}^{\tree} F_n \biggr],
\end{multline}
where the one-loop `factorisation function' $F_n$ is
helicity-independent.  Na\"ively this only contains single poles,
however for complex momenta there are double poles. These can be
interpreted as due to the three-point all-plus (or all-minus) one-loop
amplitude also containing a pole
\begin{equation}
  A^{\oneloop}_3( K^+, a^+,b^+) = { 1 \over K^2 } V_3^\oneloop(K^+,a^+,b^+)
\end{equation}
where, for pure Yang--Mills,
\begin{equation}
  V^\oneloop_3(K^+,a^+,b^+) =-{ i \over 48 \pi^2
  }\spb{K}.a\spb{a}.b\spb{b}.K   .  
\end{equation}
Explicitly, consider the amplitude~\cite{Bern:1993mq}:
\begin{multline}
  A_{5}^\oneloop (1^-, 2^+, 3^+, 4^+, 5^+) \sim {1\over \spa3.4^2}
  \biggl[ -{\spb2.5^3 \over \spb1.2 \spb5.1} \\+ {\spa1.4^3 \spb4.5
    \spa3.5 \over \spa1.2 \spa2.3 \spa4.5^2} - {\spa1.3^3 \spb3.2
    \spa4.2 \over \spa1.5 \spa5.4 \spa3.2^2} \biggr] .
\end{multline}
If we carry out a complex shift on $\lambda_5$,$\bar\lambda_1$ as in
eq.~(\ref{eqshift}) then $\spa4.5 \longrightarrow \spa4.5+z\spa4.1$
which vanishes at $z= -{\spa4.5/ \spa4.1}$ and the amplitude has a
double pole at this point. 

Computing this amplitude using $V^{\oneloop}_3$ correctly generates
the double pole in the amplitude \cite{Bern:2005hs,Brandhuber:2007up},
however it needs augmentation to give an expression with the correct
single pole. By trial and error, adding the second term in
\eqref{eqYangMillsSoft} gives the correct single pole and completes
the computation of the amplitude.

For gravity the vertex
\begin{equation}
  V^{\oneloop}(K^+,a^+,b^+) =-{ i \kappa^3(
    \spb{K}.a\spb{a}.b\spb{b}.K )^2  \over 1440 \pi^2 }
\end{equation}
can be used to generate a double pole term but
attempts~\cite{Brandhuber:2007up} to implement a universal correction
for the single pole analogous to that of \eqref{eqYangMillsSoft} have
failed.
The resolution is to replace the factorisation term of
\eqref{eq:LoopFact} with a tree insertion diagram:
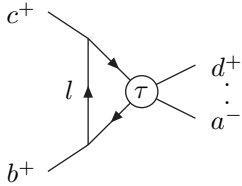
\begin{colbox}
  \begin{center}
    \begin{picture}(92,71) \SetOffset(10,-4) \ArrowLine(20,20)(20,60)
      \ArrowLine(20,60)(40,40) \ArrowLine(40,40)(20,20)
      \Line(5,10)(20,20) \Line(5,70)(20,60) \Line(40,40)(60,50)
      \Line(40,40)(60,30) \Text(-5,70)[c]{$c^+$}
      \Text(-5,10)[c]{${b}^+$} \Text(72,50)[c]{$d^+$}
      \Text(72,43)[c]{$.$} \Text(72,36)[c]{$.$}
      \Text(72,30)[c]{${a}^-$} \BCirc(40,40){6}
      \Text(40,40)[cc]{$\Tau$} \Text(13,41)[c]{$l$}
    \end{picture}
  \end{center}
\end{colbox}
which we compute using axial gauge diagrammatics. The circle in the
diagram represents the sums of all possible tree diagrams with two
internal legs and the given external legs, which we denote
$\Tau$.  Note that we evaluate these diagrams for real
momenta and only carry out analytic shifts on the final expressions.

\section{Axial gauge diagrammatics}
\label{sec:axialdiags}

Following \cite{Schwinn:2005pi} we use a set of Feynman rules for
Yang--Mills amplitudes based on scalar propagators connecting three
and four point vertices.  The starting point is the expansion of the
axial gauge propagator in terms of polarisation vectors,
\begin{equation}
  i{d_{\mu\nu}\over k^2} ={i\over k^2}
  [\epsilon_\mu^+(k)\epsilon_\nu^-(k)+\epsilon_\mu^-(k)\epsilon_\nu^+(k)+
  \epsilon_\mu^0(k)\epsilon_\nu^0(k) ],
  \label{propa}
\end{equation}
where \newcommand{\nulled}{\flat}
\begin{equation}
  \epsilon^+_\mu = {[k^\nulled | \gamma_\mu | q \ra \over \sqrt{2}
    \spa{k^\nulled}.q}, 
  \epsilon^-_\mu= { [
    q|{\gamma_\mu}|{k^\nulled} \ra \over \sqrt{2} \spb{k^\nulled}.q}, 
  \epsilon^0_\mu=2{\sqrt{k^2}\over 2k\cdot q}q_\mu,  
\end{equation}
with
\begin{equation}
  k^\nulled := k - {k^2\over 2k\cdot q} q,  \label{eqqnullification}
\end{equation}
where $q$ is a null reference momentum which may be complex. The
resulting three-point vertices are,
\begin{equation}
  \eqalign{ {1\over i\sqrt{2}}V_3^{\rm MHV}(1^-,2^-,3^+)&={\spa 1.2 {\spb 3.q}^2\over \spb 1.q \spb 2.q}, \cr
    {1\over i\sqrt{2}}V_3^{\overline{\rm MHV}}(1^+,2^+,3^-)&={\spb 2.1 {\spa 3.q}^2 \over \spa 1.q \spa 2.q}, \cr}
  \label{eqThreePointVertices}
\end{equation}
along with a $V_3(1^+,2^-,3^0)$ vertex which can be absorded into
effective four-point vertices.

When adopting a recursive approach which involves shifting a
negative-helicity leg $a$ and a positive-helicity leg $b$, the
recursion-optimised choice for the reference momentum $q$ is
\begin{equation}
  \lambda _q=\lambda _a, \qquad
  \bar\lambda _q=\bar\lambda _b.
\end{equation}
With this choice of $q$ the leg $a$ ($b$) can only enter a diagram
on a $V_3^{\rm MHV}$ ($V_3^{\overline{\rm MHV}}$) vertex,
and there are no four-point vertices in the single-minus amplitudes at
tree or one-loop level.

Singularities arise in the loop integration from the region of loop
momentum where the denominators of three adjacent propagators vanish
simultaneously. This requires the two null legs to which the
propagators connect to become collinear. In the integration region of
interest all the legs of $\Tau$ are close to null and $\Tau$
approaches the corresponding on-shell tree amplitude. The internal
legs are also close to collinear.  Helicity configurations for which
$\Tau$ is singular in this collinear limit, shown in Fig.~1, contribute to the
double (and single) pole, conversely those that give a vanishing
$\Tau$ in the collinear limit give no residue.
\begin{colbox}
  \begin{center}
    \begin{minipage}{90pt}\begin{center}
      \begin{picture}(90,72) \SetOffset(12,-3)
        \ArrowLine(20,20)(20,60) \ArrowLine(20,60)(40,40)
        \ArrowLine(40,40)(20,20) \Line(5,10)(20,20) \Line(5,70)(20,60)
        \Line(40,40)(60,50) \Line(40,40)(60,30) \Text(-5,70)[c]{$c^+$}
        \Text(-5,10)[c]{${b}^+$} \Text(72,50)[c]{$d^+$}
        \Text(72,43)[c]{$.$} \Text(72,36)[c]{$.$}
        \Text(72,30)[c]{${a}^-$} \BCirc(40,40){6}
        \Text(40,40)[cc]{$\Tau$} \Text(18,58)[tr]{\footnotesize{$-$}}
        \Text(18,22)[br]{\footnotesize{$+$}}
        \Text(22,20)[cl]{\footnotesize{$-$}}
        \Text(22,60)[cl]{\footnotesize{$+$}}
        \Text(37,48)[bc]{\footnotesize{$-$}}
        \Text(37,32)[tc]{\footnotesize{$+$}}
      \end{picture}
      \\ (a)
    \end{center}\end{minipage}\qquad
    \begin{minipage}{90pt}\begin{center}
      \begin{picture}(90,72)
        \SetOffset(12,-3)
        \ArrowLine(20,20)(20,60)
        \ArrowLine(20,60)(40,40) \ArrowLine(40,40)(20,20)
        \Line(5,10)(20,20) \Line(5,70)(20,60)
        \Line(40,40)(60,50) \Line(40,40)(60,30)     
        \Text(-5,70)[c]{$c^+$}
        \Text(-5,10)[c]{${b}^+$}
        \Text(72,50)[c]{$d^+$}
        \Text(72,43)[c]{$.$}
        \Text(72,36)[c]{$.$}
        \Text(72,30)[c]{${a}^-$}
        \BCirc(40,40){6}
        \Text(40,40)[cc]{$\Tau$}
        \Text(18,58)[tr]{\footnotesize{$+$}}
        \Text(18,22)[br]{\footnotesize{$-$}}
        \Text(22,20)[cl]{\footnotesize{$+$}}
        \Text(22,60)[cl]{\footnotesize{$-$}}
        \Text(37,48)[bc]{\footnotesize{$+$}}
        \Text(37,32)[tc]{\footnotesize{$-$}}
      \end{picture}
      \\ (b)
    \end{center}\end{minipage}
  \end{center}
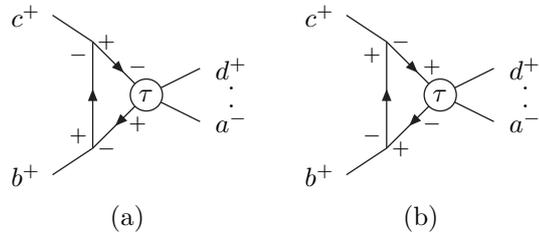
\textbf{Figure 1:} Contributing helicity structures
\end{colbox}
The diagram of Fig.~1(b) evaluates to
\begin{equation}
  \int d^4l \,{ [b|l|a\ra [c|l|a\ra \over \spa{b}.{a}\spa{c}.a } {
    \spa{C}.{a}^2 \over \spa{B}.{a}^2} { \tau(C^+,d^+,\cdots,
    a^-,B^-) \over l^2(l+k_b)^2(l-k_c)^2}
  \label{eq:diag4b}
\end{equation} 
where $B=l+b$, $C=c-l$, and the momenta in the spinor products are
$q$-nullified as in \eqref{eqqnullification}.  We construct a basis
for the loop momentum using $b$ and $c$:
\begin{equation}
  \eqalign{
    &l=\alpha_1  (k_b+k_{c})+ \alpha_2(k_b -k_{c}) +\cr
    &({\alpha_3}+{i\alpha_4}){
      \spa{c}.a \over \spa{b}.a} \lambda_b\bar\lambda_{c}+
    ({\alpha_3}-{i\alpha_4})
    { \spa{b}.a \over \spa{c}.a }
    \lambda _{c}\bar\lambda _b
    \cr}
  \label{eq:loopbasis}
\end{equation}
Under this parametrisation,
\begin{equation}
  \int { d^4l \;\;f(l)\over l^2(l+k_b)^2(l-k_c)^2} = 
  {1 \over
    s_{bc} } \int d\alpha_i \;
  F(\alpha_i)f(l(\alpha_i))
  \label{eqParaInt}
\end{equation}
where $F(\alpha_i)$ has no dependence on $s_{bc}$. The integrand from
Fig.~1(b) then becomes,
\begin{equation} {[bc]\over \la {b}c\ra}
  {\spa{C}.{a}^2 \over \spa{B}.{a}^2}
  \Tau(C^+,d^+,\dots,a^-,B^-)  \times F'(\alpha_i).
  \label{triblobYM}
\end{equation}

In order to evaluate the contribution from \eqref{triblobYM} we must
evaluate the tree structures to order $\spa{b}.c^0$. For diagrams
within $\tau$ involving $1/s_{bc}$, this means going beyond leading
order. These correspond to triangles in the full diagram and the
calculation is readily done exactly. The diagrams without this
propagator need only be calculated to leading order.  In this regard,
not only is the recursive approach selecting a subset of diagrams for
calculation, it is also allowing us to calculate these diagrams in a
very convenient limit.

For gravity the equivalent expression to \eqref{triblobYM} is
\begin{equation} {[bc]^3\over \la {b}c\ra}{\spa{C}.{a}^4 \over
    \spa{B}.{a}^4}
  \Tau_{\text{g}}(C^+,d^+,\dots,a^-,B^-)  \times \tilde F(\alpha_i).
  \label{triblob}
\end{equation}

\section{The graviton scattering amplitude
  $M^\oneloop(1^-,2^+,3^+,4^+,5^+)$}
\label{sec:grav-mpppp}
There are three types of recursive contribution to this amplitude,
which in turn are summed over the distinct permutations of $c$, $d$
and $e$.  Diagrams $R_1$ and $R_2$ involve only single poles and are
obtained from the corresponding four-point one-loop amplitudes for the
circles marked $L$.  \hfill
\newcommand{\recspA}{$\Rec_1$} \newcommand{\recspB}{$\Rec_2$}
\newcommand{\recdp}{$\Rec_3$}
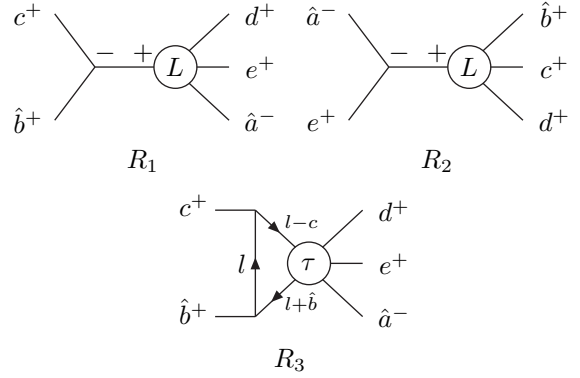
\begin{colbox}
  \begin{center}
    \begin{minipage}{100pt}\begin{center}
        \begin{picture}(100,52)
          \SetOffset(22,-14)
          \CArc(40,40)(8,0,360)
          \Line(-5,20)(10,40) \Line(-5,60)(10,40)
          \Line(10,40)(32,40)
          \Line(45,46)(60,60) \Line(45,34)(60,20) \Line(48,40)(60,40)
          \Text(-15,60)[c]{$c^+$} \Text(40,40)[c]{$L$} \Text(-15,20)[c]{${\hat
              b}^+$} \Text(72,60)[c]{$d^+$} \Text(72,40)[c]{$e^+$}
          \Text(72,20)[c]{${\hat a}^-$}
          \Text(10,42)[bl]{$-$}
          \Text(32,42)[br]{$+$}
        \end{picture}
        \\ \recspA\end{center}\end{minipage}\quad
    \begin{minipage}{100pt}\begin{center}
        \begin{picture}(100,52)
          \SetOffset(22,-14)
          \CArc(40,40)(8,0,360)
          \Line(-5,20)(10,40) \Line(-5,60)(10,40) \Line(10,40)(32,40)
          \Line(45,46)(60,60) \Line(45,34)(60,20) \Line(48,40)(60,40)
          \Text(-15,60)[c]{$\hat a^-$} \Text(40,40)[c]{$L$}
          \Text(-15,20)[c]{${e}^+$} \Text(72,60)[c]{${\hat b}^+$}
          \Text(72,40)[c]{$c^+$} \Text(72,20)[c]{${d}^+$}
          \Text(10,42)[bl]{$-$}
          \Text(32,42)[br]{$+$}
        \end{picture}
        \\ \recspB\end{center}\end{minipage} \vskip 2ex
    \begin{minipage}{100pt}\begin{center}
        \begin{picture}(90,52)
          \SetOffset(12,-13)
          \ArrowLine(20,20)(20,60)
          \ArrowLine(20,60)(35,46) \ArrowLine(35,34)(20,20)
          \CArc(40,40)(8,0,360) \Line(5,20)(20,20) \Line(5,60)(20,60)
          \Line(45,46)(60,60) \Line(45,34)(60,20) \Line(48,40)(60,40)
          \Text(40,40)[c]{$\Tau$} \Text(31,24)[l]{${}^{l+{\hat b}}$}
          \Text(31,54)[l]{${}^{l-c}$} \Text(15,40)[c]{$l$}
          \Text(-3,62)[c]{$c^+$} \Text(-3,22)[c]{${\hat b}^+$}
          \Text(72,60)[c]{$d^+$} \Text(72,40)[c]{$e^+$}
          \Text(72,20)[c]{${\hat a}^-$}
        \end{picture}
        \\ \recdp\end{center}\end{minipage}
  \end{center}
\end{colbox}

Doing recursion with the shift \eqref{eqshift}, we obtain
\begin{align}
  \begin{split}
    &\Rec_1(a,b,c,d,e)={1 \over 5760}{\la ad\ra^2 \la ae\ra^2
      [bc][de]^4 \over \la ab\ra^2\la bc\ra\la ce\ra^2\la cd\ra^2\la
      de\ra^2}\times \\&\quad \left( \la cd\ra^2 \la ae\ra^2+\la
      ac\ra\la cd\ra\la de\ra\la ae\ra+\la ac\ra^2\la de\ra^2 \right),
  \end{split}\\
  \begin{split}
    &\Rec_2(a,b,c,d,e)=-{3\over 5760}{\la ae\ra [be]^4 \over \la
      cd\ra^2 [ab]^2 [ae]} \times \\&\quad \left([bc]^2[ de]^2+[ bc][
      cd][ de][ be]+[ cd]^2[ be]^2\right). \label{eq:R1and2}
  \end{split}
\end{align}

Diagram \recdp\ contains a double pole so we must evaluate
$\Tau_{\text{g}}$ of \eqref{triblob}.  We use the five-point KLT
relation~\cite{Kawai:1985xq},
\begin{gather}\begin{split}
    &M(a^-B^-C^+d^+e^+) =\\ &\quad s_{BC}s_{de} A(a^-B^-C^+d^+e^+)
    A(a^-C^+B^-e^+d^+)+ \\ & s_{Bd} s_{Ce}
    A(a^-B^-d^+C^+e^+)A(a^-d^+B^-e^+C^+)
    \label{eqKLT5pt}
  \end{split}\end{gather}
in a form that restricts the $\langle b \, c \rangle$ pole to the
first term.
We calculate this as Laurent series in $\langle b\:c \rangle$,
dropping terms that will not contribute to the residues.  While the
KLT relations are only valid for on-shell momenta, we assume the
deviation of \eqref{eqKLT5pt} from a direct off-shell calculation may
be neglected \footnote{The general case is worthy of further
  study~\cite{Alstonetal}.} in the region around $B^2=C^2=0$.
 
In our choice of axial gauge, $A(a^-B^-C^+d^+e^+)$ receives
contributions from five diagrams, only two of which contain a
$V_3(B^-,C^+,x)$ vertex and thus contribute to $\Tau$'s collinear
singularity. Denoting these by $D_a$ and $D_b$, we find, using
\eqref{eq:loopbasis}
\begin{multline}
  D_a+D_b= {\la Ba\ra^2\over\la Ca\ra^2} {\la a|bc|a\ra \over
    s_{bc} [ab]\la da\ra\la ea\ra} \times \\
  \biggl( {[b|ad|e]-[b|cb|e]\over [ae]\la de\ra} \biggr)
  f_a(\alpha_i),
  \label{BCfuse}
\end{multline}
where $f_a(\alpha_i)$ is some function that depends only on the
integral parameters, $\alpha_i$. We note that the second term is
sub-leading in the $\la b\,c\ra$ pole.

The leading pole in $A(a^-C^+B^-e^+d^+)$ is obtained similarly and we
obtain the full leading $\langle b\,c\rangle$ pole in \eqref{eqKLT5pt}
as
\begin{equation}
  {\la Ba\ra^4\over\la Ca\ra^4} s_{bc}s_{de}{\la ab\ra^2
    \la ac\ra^2 [de]^3 [bc] \over \la bc\ra \la de\ra
    [ a|d+e|a \ra
  }
  f_a'(\alpha_i). 
\end{equation}
\def\dpf{{\cal D}}%
Combining this with the factors arising from the
left hand part of the full diagram (\cf\ \eqref{triblob}) and
integrating over the $\alpha_i$ the leading term in the Laurent series
for \recdp\ is proportional to
\begin{equation} {[bc]^4 \la ab\ra^2 \la ac\ra^2 [de]^3 \over \la
    bc\ra^2 \la de\ra[ a|d+e|a \ra }
  \equiv \dpf
  , 
  \label{eq:Rec3}
\end{equation}
which clearly displays the double pole factor.

We now express each sub-leading contribution to \eqref{triblob} as
$\dpf\times\delta_j f_j(\alpha_i)$.  Firstly there is the
sub-leading contribution of \eqref{BCfuse} together with the
corresponding contribution from $A(a^-C^+B^-e^+d^+)$:
\begin{equation}
  \delta_1 = {s_{bc}[be]\over [b|ad|e]}+ {s_{bc}[bd]\over [b|ae|d]}.
\end{equation}
The remaining diagrams for $A(a^-B^-C^+d^+e^+)$ (and its counterpart
$A(a^-C^+B^-e^+d^+)$), in which $B$ and $C$ enter on different
vertices contribute
\begin{equation}
  \delta_2={s_{bc}[e|a|c\ra \over s_{ab}[e|d|c\ra },
\end{equation}
\begin{equation}
  \delta_3=
  {\la bc\ra \la de\ra \over s_{ab}[de]}\biggl( {[e|B|a\ra
    [eb] \over \la da\ra \la cd\ra}+{[d|B|a\ra [db] \over \la ea\ra \la
    ce\ra} \biggr).
\end{equation}
These diagrams are finite in the collinear limit, so we can drop terms
proportional to $B^2$ and $C^2$.  Finally we need the second term in
\eqref{eqKLT5pt}, which is also finite in the collinear limit and can
be evaluated using MHV tree amplitudes, yielding:
\begin{equation}
  \delta_4 ={\la bc\ra\la de\ra [d|B|a\ra[e|C|a\ra \over [bc][de]\la
    ab\ra^2\la cd\ra\la ce\ra}.
\end{equation}

Up to $\mathcal{O}(\langle b\,c\rangle^{-1})$ \eqref{triblob} is then
expressed as
\begin{equation} {[bc]^4 \la ab\ra^2 \la ac\ra^2 [de]^3 \over \la
    bc\ra^2 \la de\ra[ a|d+e|a \ra }
  \biggl( 1 +\sum_j \delta_j f_j(\alpha_i)\biggr)F'(\alpha_i).
  \label{eq:allorderR3}
\end{equation}
This has purely polynomial dependence on the $\alpha_i$. The
integration thus gives constant numerical factors which may be
obtained by direct evaluation or, more conveniently, by considering
collinear limits.

We now determine the amplitude recursively by applying the shift
\eqref{eqshift} to the integrated \eqref{eq:allorderR3} and evaluating
the residue at $z=-{\la bc\ra/\la ac\ra}$.  The coefficient of the
double pole has a $z$ dependence under this shift which generates a
further contribution to the single pole since
\begin{equation}
  \Res_{z=z_i}  { f(z) \over z(z-z_i)^2 }= -{ f(z_i)\over
    z_i^2 } + { 1 \over z_i } \left.\frac{df}{dz}\right|_{z=z_i}. 
\end{equation}
The full contribution from \recdp\ is then
\begin{multline}
  \Rec_3(a,b,c,d,e)={1 \over 5760} {\la ab\ra^2\la ac\ra^4[bc]^4[de]
    \over\la ad\ra\la ae\ra\la bc\ra^2\la cd\ra\la ce\ra\la de\ra}\\
  \times \bigl(1+\Delta(a,b,c,d,e)\bigr),
\end{multline}
where
\begin{multline}
  \Delta(a,b,c,d,e) = -{1\over2}{\la ad\ra\la bc\ra\over \la ab\ra\la
    cd\ra} -{1\over2}{\la ae\ra\la bc\ra\over \la ab\ra\la
    ce\ra} \\
  -3 {[db][eb]\la bc\ra \la de\ra\over \la dc\ra\la ec\ra [bc][de]} -3
  {[dc][ec] \la bc\ra\la de\ra\la ca\ra^2 \over \la dc\ra\la ec\ra
    [bc] [de] \la ba\ra^2}
  \\
  -{7\over2} {[dc][eb]\la bc\ra \la de\ra\la ca\ra \over\la dc\ra\la
    ec\ra [bc] [de] \la ba\ra} -{7\over2} {[db][ec]\la bc\ra\la
    de\ra\la ca\ra \over\la dc\ra\la ec\ra [bc][de]\la ba\ra}.
\end{multline}

The full amplitude is the sum over contributions arising from three
orderings of external legs,
\begin{multline}
  M^\oneloop(1^-,2^+,3^+,4^+,5^+) = \Rec(1,2,3,4,5) +\\
  \Rec(1,2,4,5,3) + \Rec(1,2,5,3,4),
  \label{eq:recursion-amp}
\end{multline}
(the full amplitude has a factor of $i\kappa^5/16\pi^2$), and each
$\Rec$ is the sum of the recursive diagrams,
\begin{equation}
  \Rec=\Rec_1+\Rec_2+\Rec_3.
\end{equation}
This expresion has the correct collinear limits, is symmetric under
interchange of pairs of positive-helicity legs and agrees numerically
with that calculated by string-based rules.  We have also calculated
$M^\oneloop(1^-,2^+,3^+,4^+,5^+,6^+)$~\cite{Dunbar:2010xk}, and again
checked that it has the correct symmetries and collinear limits.
\textit{Mathematica} code for the five- and six-point amplitudes may
be found at \url{http://pyweb.swan.ac.uk/~dunbar/graviton.html}.

\section{Conclusions and remarks}
We have demonstrated how to augment recursion to determine the
rational terms in amplitudes with double poles under a complex
shift. Double poles are unavoidable in the case of the amplitudes
$A^\oneloop(1^-,2^+,3^+,\ldots ,n^+)$ in both Yang-Mills and gravity.
In the absence of a universal soft factor analogous to
\eqref{eqYangMillsSoft}, to perform the augmented recursion the
sub-leading poles must be determined on a case-by-case basis. While we
have done this for both the five- and six-point single-minus gravity
amplitudes, this procedure could be used to calculate any higher-point
single-minus amplitude.

\end{document}